\documentclass[final,5p,times,twocolumn]{elsarticle}

\usepackage[T1]{fontenc}
\usepackage[utf8]{inputenc}

\usepackage{amsmath}
\usepackage{amsfonts}
\usepackage{amssymb}
\usepackage{amsxtra}
\usepackage{array}
\usepackage{graphicx}
\usepackage{hepunits}
\usepackage{units}
\usepackage{color}
\usepackage{tikz}
\usepackage[linkcolor=blue,citecolor=blue,urlcolor=blue,colorlinks=true,breaklinks]{hyperref}
\usepackage{xspace}
\usepackage[normalem]{ulem}

\usepackage[mathscr,scaled=1.15]{urwchancal}

\newcommand{\be}{\begin{equation}}
\newcommand{\bea}{\begin{eqnarray}}
\newcommand{\ee}{\end{equation}}
\newcommand{\eea}{\end{eqnarray}}
\def\1eq#1{Eq.~(\ref{#1})}

\def\2eqs#1#2{Eqs.~(\ref{#1}) and~(\ref{#2})}
\def\3eqs#1#2#3{Eqs.~(\ref{#1}),~(\ref{#2}) and~(\ref{#3})}
\def\tlambda{\mkern 2mu\tilde{\mkern-4mu \lambda \mkern-2mu}\mkern 1.2mu}  %lambda tilde.
\newcommand{\ie}{\textit{i.e. }}
\newcommand{\eg}{\textit{e.g. }}

\newcommand{\fatg}{{\rm{I}}\!\Gamma}

\biboptions{sort&compress}

\journal{Physics Letters B}

\begin{document}

\begin{frontmatter}

\title{Lattice three-gluon vertex in extended kinematics: planar degeneracy}

\author[UPO]{F.~Pinto-G\'omez}
%\author[UNICAMP]{A.~C.~Aguilar}
\author[UPO]{F.~De Soto}
\author[UNICAMP]{M.~N. Ferreira}
\author[UV]{J.~Papavassiliou}
\author[Huelva]{J.~Rodr\'iguez-Quintero}
\address[UPO]{Dpto. Sistemas F\'isicos, Qu\'imicos y Naturales, Univ. Pablo de Olavide, 41013 Sevilla, Spain}
\address[UNICAMP]{\mbox{University of Campinas - UNICAMP, Institute of Physics ``Gleb Wataghin,''} 
13083-859 Campinas, S\~{a}o Paulo, Brazil}
\address[UV]{Department of Theoretical Physics and IFIC, University of Valencia and CSIC, E-46100, Valencia, Spain}
\address[Huelva]{Dpto. Ciencias Integradas, Centro de Estudios Avanzados en Fis., Mat. y Comp., Fac. Ciencias Experimentales, Universidad de Huelva, Huelva 21071, Spain}

\begin{abstract}
We present novel results 
for the three-gluon vertex, obtained from 
an extensive quenched lattice simulation in the Landau gauge. 
The simulation evaluates the transversely projected vertex, spanned on a special tensorial basis, whose form factors are naturally parametrized in terms of individually Bose-symmetric variables.
Quite interestingly, when evaluated 
in these kinematics, the corresponding form factors depend 
almost exclusively on a single kinematic variable, 
formed by the sum of the squares of the three incoming four-momenta, $q$, $r$, and $p$.  Thus, 
all configurations lying on a given plane 
in the coordinate system $(q^2, r^2, p^2)$ 
share, to a high degree of accuracy, the same form factors, a property that we 
denominate \emph{planar degeneracy}.
We have confirmed the validity of this property through an exhaustive study of 
the set of configurations satisfying the condition $q^2 = r^2$, within the range 
$[0, 5\, \rm GeV]$.
Moreover, a preliminary exploration
reveals that the planar degeneracy 
persist in the case of 
more arbitrary configurations. 
This drastic simplification allows for a remarkably compact description of the main bulk of the data, which is particularly suitable for future numerical applications. 
 A semi-perturbative analysis reproduces the lattice findings  
 rather accurately,  
 once the inclusion of a gluon mass has cured  
 all spurious divergences.

\end{abstract}


\end{frontmatter}

\noindent\textbf{1.$\;$Introduction}. 
%\section{Introduction}
%
The three-gluon vertex plays a central role in the intricate infrared dynamics of Quantum Chromodynamics (QCD)\,\cite{Marciano:1977su,Ball:1980ax,Davydychev:1996pb},
and the detailed exploration of its salient nonperturbative features has attracted particular attention in recent years\,\cite{Alkofer:2004it,Cucchieri:2006tf,Cucchieri:2008qm,Huber:2012zj,Pelaez:2013cpa,Aguilar:2013vaa,Blum:2014gna,Eichmann:2014xya,Mitter:2014wpa,Williams:2015cvx,Blum:2015lsa,Cyrol:2016tym,Athenodorou:2016oyh,Duarte:2016ieu,Boucaud:2017obn,Aguilar:2019uob,Aguilar:2021lke,Aguilar:2021okw,Catumba:2021yly,Catumba:2021hng,Sternbeck:2017ntv,Corell:2018yil,Aguilar:2019jsj,Aguilar:2019kxz,Vujinovic:2018nqc,Barrios:2022hzr}. 
This ongoing search, based on the profitable synergy between lattice simulations and continuum methods,
has afforded a firmer grip on delicate underlying patterns,  
establishing prominent connections with the emergence of a mass scale in the gauge sector of the theory\,\cite{Aguilar:2008xm,Boucaud:2008ky,Fischer:2008uz,Dudal:2008sp,Tissier:2010ts,Cloet:2013jya,Pelaez:2014mxa,Eichmann:2021zuv,Gao:2017uox,Roberts:2021xnz,Binosi:2022djx,Roberts:2020udq,Roberts:2021nhw,Papavassiliou:2022wrb,Roberts:2020hiw}.  
In addition to its theoretical importance, the three-gluon vertex is a central component in a variety of phenomenological 
studies in the continuum. In particular, 
the outstanding feature of {\it infrared suppression}\,\cite{Cucchieri:2006tf,Cucchieri:2008qm,Huber:2012zj,Pelaez:2013cpa,Aguilar:2013vaa,Blum:2014gna,Eichmann:2014xya,Mitter:2014wpa,Williams:2015cvx,Blum:2015lsa,Cyrol:2016tym,Athenodorou:2016oyh,Duarte:2016ieu,Corell:2018yil,Boucaud:2017obn,Aguilar:2019jsj,Aguilar:2019uob,Aguilar:2019kxz,Aguilar:2021lke,Aguilar:2021okw,Catumba:2021yly,Catumba:2021hng,Sternbeck:2017ntv} displayed by 
its main form factors is instrumental for
the formation of bound states with the right 
physical properties\,\cite{Meyers:2012ka,Binosi:2014aea,Souza:2019ylx,Binosi:2016nme,Roberts:2020hiw,Huber:2018ned,Athenodorou:2020ani,Athenodorou:2021qvs,Huber:2021yfy}. 

If we denote by $q$, $r$, and $p$, the three four-momenta entering into the three-gluon vertex, with $q+r+p=0$, 
the corresponding form factors are functions of $q^2$, $r^2$, and $p^2$, or, equivalently,
$q^2$, $r^2$, and the angle $\theta_{qr}$ formed between $q$ and $r$.
However, the analysis of lattice simulations in general kinematics is particularly 
costly. Consequently, to date, SU(3) lattice studies have been restricted mainly to special kinematics involving a single
momentum scale, such as the  
``{\it symmetric}\,'' ($q^2 = r^2=p^2$) and the ``{\it soft-gluon}\," ($q^2=r^2$, $\theta_{qr} =\pi$) configurations\,\cite{Athenodorou:2016oyh,Duarte:2016ieu,Boucaud:2017obn,Aguilar:2019uob,Aguilar:2021lke,Aguilar:2021okw,Catumba:2021yly,Catumba:2021hng}. In more general kinematics 
only very preliminary results are available
\,\cite{Sternbeck:2017ntv}, or they are specialized to the SU(2) gauge group\,\cite{Cucchieri:2006tf,Cucchieri:2008qm}. 

Even though plenty has already been learned from the aforementioned special configurations\,\cite{Athenodorou:2016oyh,Duarte:2016ieu,Boucaud:2017obn,Aguilar:2019uob,Aguilar:2021lke,Aguilar:2021okw,Catumba:2021yly,Catumba:2021hng,Sternbeck:2017ntv,Corell:2018yil,Aguilar:2019jsj,Aguilar:2019kxz}, 
it would be clearly advantageous to acquire lattice data for
the pertinent form factors in more general kinematics.
Such novel information would help us with the 
systematic refinement of continuum approaches, and could be 
decisive in validating the dynamical picture of 
gluon mass generation through 
the operation of the Schwinger mechanism in QCD\,\cite{Schwinger:1962tn,Schwinger:1962tp,Cornwall:1981zr,Bernard:1982my,Donoghue:1983fy,Poggio:1974qs,Smit:1974je,Wilson:1994fk,Philipsen:2001ip,Aguilar:2011ux,Aguilar:2015bud,Papavassiliou:2022wrb}.

In the present work we carry out a lattice simulation of the
transversely projected three-gluon vertex, denoted by 
$\overline{\Gamma}_{\alpha \mu \nu}(q,r,p)$, 
using quenched SU(3) field configurations in the Landau gauge.
$\overline{\Gamma}_{\alpha \mu \nu}(q,r,p)$ is expanded on a special basis comprised by four fully transverse 
and {\it individually} Bose-symmetric tensors.
Consequently, the 
corresponding form factors must be functions of Bose symmetric combinations of the kinematic variables, the most relevant 
being $s^2 = \frac{1}{2}(q^2+r^2+p^2)$, representing a plane
in the coordinate system $(q^2, r^2, p^2)$. 

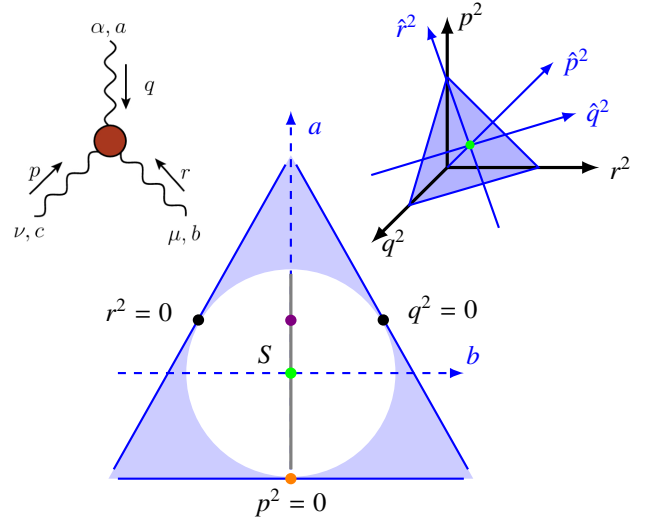
\begin{figure}[t]
\begin{tabular}{ccc}
\begin{tabular}{c}
\includegraphics[scale=0.3]{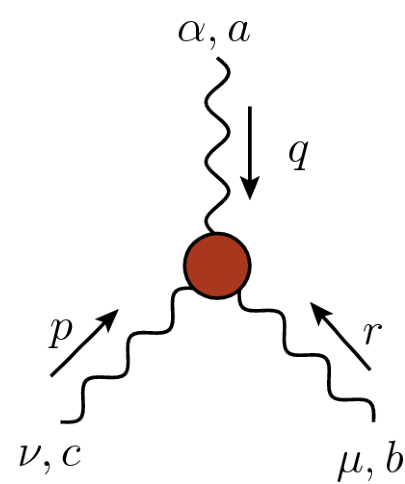} 
		\\
		\rule[0cm]{0cm}{3.5cm}	
\end{tabular}			
&		
\hspace*{-2.3cm}
\begin{tabular}{c}
\rule[0cm]{0cm}{1cm} \\
		\begin{tikzpicture}[scale=0.8]
			% NODES
			\node (r) at ( 3.0,  0.0) {}; % 
			\node (q) at (-3.0,  0.0) {}; % 
			\node (p) at ( 0.0, 5.36) {}; % 
			\node (p2) at ( 0.0, 3.55) {}; 
			\node (S) at (0,1.75) {};
			\node (Op) at (0,2.625) {};
			\node (Oq) at (-0.76,1.31) {};
			\node (Or) at (0.76,1.31) {};
			\node (SGp) at (0,0) {};
			\node (SGq) at (1.52,2.63) {};
			\node (SGr) at (-1.52,2.63) {};
			\node (apA) at ( 3.0,  1.75) {}; % 
			\node (amA) at ( -3.0,  1.75) {}; % 
			\node at (3.0,2.05) [blue] {$b$};
			\node (apB) at ( 0,  6.25) {};
			\node at (0.4,5.8) [blue] {$a$};
			%fillings
			\fill[fill=blue!20] (p.center) -- (r.center) -- (q.center) -- (p.center);	-
%			\node at (3.5,0) {$r^2$}; 
%			\node at (-3.5,0) {$q^2$}; 
%			\node at (-0.5,5.6) {$p^2$}; 
			\draw [thick, blue] (r) -- (q) -- (p) -- (r);
%			\draw [thick, red] (S) circle (1.75cm);
			\fill[fill=white] (S) circle (1.72cm);
			\draw [thick,dashed, blue, -latex] (SGp) -- (apB);
			\draw [thick,dashed,blue, -latex] (amA) -- (apA);
			% temp to locate points
			%\draw [thick, red] (S) circle (0.875cm);
			%\draw [gray] (r) -- (SGr);
			%\draw [gray] (p) -- (SGp);
			\draw [very thick,gray] (p2) -- (SGp);
			% node points
			\node at (-0.4,2.05) {$S$};
			\node at (S)[circle,fill,inner sep=1.5pt,green]{};
			%\node at (0.4,2.7) {$O_p$};
			\node at (Op) [circle,fill,inner sep=1.5pt,violet]{};
			%\node at (-1.1,1.5) {$O_q$};
			%\node at (Oq) [circle,fill,inner sep=1.5pt]{};
			%\node at (+1.1,1.5) {$O_r$};
			%\node at (Or) [circle,fill,inner sep=1.5pt]{};
			\node at (0,-0.4) {${p^2=0}$};
			\node at (SGp) [circle,fill,inner sep=1.5pt,orange]{};
			\node at (2.5,2.8) {${q^2=0}$};
			\node at (SGq) [circle,fill,inner sep=1.5pt,black]{};
			\node at (-2.5,2.8) {${r^2=0}$};
			\node at (SGr) [circle,fill,inner sep=1.5pt,black]{};
		\end{tikzpicture} 
\end{tabular}
&
\hspace*{-2.5cm}
\begin{tabular}{c}
		\begin{tikzpicture}[scale=0.4]
			\fill[fill=blue!30] (0,3) -- (3,0) -- (-1.25,-1.25) -- (0,3);
			\draw [very thick, -latex] (0,0) -- (0,5) node [right] {$p^2$};
			\draw [very thick, -latex] (0,0) -- (-2.5,-2.5) node [right] {$q^2$};
			\draw [very thick, -latex] (0,0) -- (5,0) node [right] {$r^2$};
			\draw [thick, blue] (0,3) -- (3,0) -- (-1.25,-1.25) -- (0,3);
			\draw [thick, blue, -latex] (0,0) -- (3.5,3.5) node [right]{$\hat{p}^2$};
			\draw [thick, blue, -latex] (1.7,-2.0) -- (-0.65,4.7) node [left] {$\hat{r}^2$};
			\draw [thick, blue, -latex] (-2.5,-0.25) -- (4.25,1.8) node [right] {$\hat{q}^2$};
			\node at (0.75,0.75) [circle,fill,inner sep=1.2pt,green]{};
		\end{tikzpicture}
		\\
		\rule[0cm]{0cm}{3.5cm}		
\end{tabular}		 
\end{tabular}
%\vspace{-0.5cm}
\caption{The kinematic configuration of the three-gluon vertex in Eq.\,\eqref{eq:Green3g}, (left diagram) represented by the Cartesian coordinates $(q^2,r^2,p^2)$, rotated to $(\hat{q}^2,\hat{r}^2,\hat{p}^2)$, according to Eqs.\,\eqref{eq:hatqrp} (right picture). Properly rescaled, such that planar coordinates are $a$=$\hat{r}^2/[\sqrt{6} \hat{p}^2]$ and $b$=$\hat{q}^2/[\sqrt{6} \hat{p}^2]$, they define an unitary equilateral triangle, whose incircle contains the class of configurations having the same angles. The bisectoral line (thick gray), and the particular soft-gluon (orange solid circle), symmetric (green) and $p^2$=$2q^2$=$2r^2$ (violet) cases appear depicted. The other two soft-gluon limits (black) are also illustrated.  
}
\label{fig:triangle}
\end{figure}

Our analysis is mostly restricted to   
kinematic configurations 
that satisfy the condition $q^2=r^2$; due to their characteristic 
geometrical representation (see Fig.~\ref{fig:triangle}), they 
are dubbed ``{\it bisectoral}\,''.\,
The results obtained reveal a rather striking pattern: 
the form factors depend almost exclusively on a single variable, namely $s_b^2 = q^2+p^2/2$, which is simply the $s^2$ introduced above evaluated at $q^2=r^2$. 
In fact, an exploratory study away from the 
bisectoral kinematics 
indicates the persistence of this singular feature: 
the form factors whose $s^2$ variable lies on a given   
plane may be very accurately described by a common 
set of form factors. In what follows we 
refer to this property as ``planar degeneracy". 
It is important to mention that this particular 
pattern was first identified in  
the continuum analysis of~\cite{Eichmann:2014xya}. 

Prompted by these key observations, 
a very simple formula [{\it viz.} \1eq{eq:compact}] 
is proposed, which enables a faithful     
description of the entire range of 
bisectoral kinematics. The  
single dynamical component of this formula 
is the soft-gluon form factor, 
whose behaviour in a wide range of momenta 
is very well understood\,\cite{Aguilar:2019uob,Aguilar:2021lke,Aguilar:2021okw}. Such a compact description 
affords considerable simplifications to a variety of 
situations where the three-gluon vertex 
must be included in the dynamical analysis. 

Finally, in order to acquire an analytic grasp  
on the observed patterns, we compute the dominant  
form factor from the corresponding 
one-loop Feynman diagrams. 
It turns out that the results are plagued by 
collinear divergences, which completely distort 
any signal of planar degeneracy. However, 
once the gluon propagator employed has been supplemented  
with an effective mass, in conformity with its well established infrared saturation\,\cite{Cucchieri:2007md,Bogolubsky:2007ud,Bogolubsky:2009dc,Oliveira:2009eh,Ayala:2012pb,Aguilar:2004sw,Aguilar:2006gr,Aguilar:2008xm,Boucaud:2008ky,Fischer:2008uz,Dudal:2008sp,RodriguezQuintero:2010wy,Tissier:2010ts,Cucchieri:2011ig,Pennington:2011xs,Cloet:2013jya,Fister:2013bh,Pelaez:2014mxa,Cyrol:2014kca,Cyrol:2018xeq}, one clearly observes the 
restoration of this property at a high degree of accuracy. 

\smallskip

\noindent\textbf{2.$\;$Kinematic configurations}. 
%\section{Kinematic configurations}
%
The starting point of 
our investigation is the quantity 
\be
{\cal G}_{\alpha \mu \nu}(q,r,p) \ = \  \frac 1 {24} f^{abc} \langle \widetilde{A}^a_\alpha(q) \widetilde{A}^b_\mu(r) \widetilde{A}^c_\nu(p) \rangle  \,,  \label{eq:Green3g}
\ee 
where $\langle \widetilde{A}^a_\alpha(q) \widetilde{A}^b_\mu(r) \widetilde{A}^c_\nu(p) \rangle$ denotes 
the three-point correlation function in Fourier space, composed by SU(3) gauge fields, $\widetilde{A}^a_\alpha$, evaluated at the four-momenta $q$, $r$ and $p$ (see Fig.\,\ref{fig:triangle}). 
Note that the above definition projects out the 
color structure of the three-point function 
proportional to $f^{abc}$ (with $f^{abc}f^{abc} =24$), 
annihilating completely any contribution proportional to the fully symmetric $d^{abc}$.

In general, an arbitrary kinematic configuration is described 
in terms of the three squared momenta, $q^2$, $r^2$, and $p^2$ .
Equivalently, one may choose 
any two of the squared momenta
and the angle spanned between them, \eg,
$q^2$, $r^2$, and $\theta_{qr}$, with 
$\cos{\theta_{qr}} =  (p^2-q^2-r^2)/2\sqrt{q^2 r^2}$;\, 
completely analogous expressions hold for $\theta_{rp}$ and $\theta_{pq}$.

A more symmetric description of the three-gluon kinematics   
arises from the properties of the 
irreducible representations of the permutation group
\,\cite{Eichmann:2014xya}. 
Its simple geometric  
derivation is obtained by noting that,   
any configuration described  
by the Cartesian coordinates $(q^2,r^2,p^2)$ 
may be rotated into 
$(\hat{q}^2,\hat{r}^2,\hat{p}^2)$,
with 
%---
\begin{subequations}
\label{eq:hatqrp}
\begin{align} 
\label{eq:hatq}
\hat{q}^2 &=\left( r^2 - q^2 \right) / \sqrt{2} \,, \\ \label{eq:hatr}
\hat{r}^2 &=  \left( 2p^2 - q^2 - r^2 \right) / \sqrt{6} \;,  \\ \label{eq:hatp}
\hat{p}^2 &= \left(q^2+r^2+p^2\right) / \sqrt{3} \;.
\end{align}
\end{subequations}
The coordinate $\hat{p}^2$ expresses the distance along the octant diagonal, while $\hat{q}^2$ and $\hat{r}^2$ locate the position on its perpendicular plane, as shown in Fig.\,\ref{fig:triangle}. This plane defines an equilateral triangle of side $\sqrt{6} \hat{p}^2$ within the positive octant. The points contained  
in the incircle of this triangle 
satisfy 
\begin{equation}
\hat{q}^4 + \hat{r}^4 \leq \frac 1 2 \hat{p}^4  \;;
\end{equation}
this relation is obtained from 
momentum conservation, and displays the kinematically allowed 
domain: a cone around the $\hat{p}^2$-axis. 

Let us next consider the class of 
kinematic configurations that 
share the same values for their angles (\ie with common $\theta_{qp}=c_1$ and 
$\theta_{rp}=c_2$).
It turns out that 
every such class has a unique representative within the incircle of a unitary equilateral triangle, as in Fig.\,\ref{fig:triangle}. Conversely, every point in the incircle of this triangle is the unique representative of a such a class. 

In particular, the bisectoral kinematics 
form a special ensemble of such configurations, 
defined by $\theta_{qp}=\theta_{rp}=c_1$, 
$\forall c_1\in [0, \pi\,]$; 
as the name suggests, all representative 
points lie on the bisectoral line drawn in Fig.\,\ref{fig:triangle} (thick gray line). 
Some special cases of bisectoral kinematics 
that will be discussed below are also illustrated:  the \emph{soft-gluon}, $c_1 = \pi/2$ (orange points); the \emph{symmetric}, $c_1 = 2\pi/3$ (green point); and the case $p^2$=$2q^2$=$2r^2$, $c_1=3\pi/4$ (violet point).       

\smallskip

\noindent\textbf{3.$\;$Transverse basis with Bose-symmetric form factors}. 
%\section{Transverse basis with Bose-symmetric form factors}
%
The connection between the ${\cal G}_{\alpha \mu \nu}(q,r,p)$ defined  in Eq.\,\eqref{eq:Green3g} and the usual one-particle irreducible (1PI) three-gluon vertex function becomes manifest by setting 
\be\label{eq:gGD3}
{\cal G}_{\alpha \mu \nu}(q,r,p) = g \overline{\Gamma}_{\alpha \mu \nu}(q,r,p) \Delta(q^2) \Delta(r^2) \Delta(p^2)  \ , 
\ee
with
\begin{subequations} 
\label{eq:G2andG3}
\begin{align}
\label{eq:barGamma}
\overline{\Gamma}_{\alpha \mu \nu}(q,r,p) &=  \fatg^{\alpha' \mu' \nu'}(q,r,p) P_{\alpha' \alpha}(q) P_{\mu' \mu}(r) P_{\nu' \nu}(p)  \, , \, \\ 
\label{eq:Green2g}
\Delta(p^2) &= \frac 1 {24} \delta^{ab} P_{\mu\nu}(p)  
\langle \widetilde{A}^a_\mu(p) \widetilde{A}^b_\mu(-p) \rangle \;;
\end{align}
\end{subequations}
where $\overline{\Gamma}$ denotes the 
{\it transversely projected} vertex \,\cite{Eichmann:2014xya,Aguilar:2019uob,Aguilar:2019kxz}
while $\fatg$ is the 1PI vertex shown schematically in Fig.\,\ref{fig:triangle}.  
In addition, $g$ is the gauge coupling, $\Delta(q^2)$ the scalar component of the gluon propagator,  obtained from the corresponding two-point correlation function; and $P_{\mu\nu}(p)=g_{\mu\nu} - p_\mu p_\nu/p^2$ stands for the standard transverse projector. Evidently, $q^\alpha {\cal G}_{\alpha\mu\nu} = r^\mu {\cal G}_{\alpha\mu\nu} = p^\nu {\cal G}_{\alpha\mu\nu}=0$. 

Note that the vertex $\fatg^{\alpha\mu\nu}$ 
consists of a pole-free component, to be denoted by 
$\Gamma^{\alpha\mu\nu}$,  and a term $V^{\alpha\mu\nu}$ 
comprised by longitudinally coupled massless poles of the type
$q^{\alpha}/q^2$, $r^{\mu}/r^2$, and $p^{\nu}/p^2$\,\cite{Aguilar:2011xe,Ibanez:2012zk,Binosi:2017rwj}.  $V^{\alpha\mu\nu}$ 
triggers the Schwinger mechanism\,\cite{Schwinger:1962tn,Schwinger:1962tp,Jackiw:1973tr,Eichten:1974et} but 
drops out from the r.h.s. of  
\1eq{eq:barGamma}, where only $\Gamma^{\alpha\mu\nu}$ contributes.

Next, recall that $\Gamma^{\alpha\mu\nu}(q,r,p)$
is usually written in the standard Ball-Chiu (BC) basis~\cite{Ball:1980ax,Davydychev:1996pb}, according to 
\begin{align}
\Gamma^{\alpha\mu\nu}(q,r,p) &= \sum_{i=1}^{10} X_i(q^2,r^2,p^2) \, \ell_i^{\alpha\mu\nu}(q,r,p) \nonumber \\ 
&+ \sum_{i=1}^4 Y_i(q^2,r^2,p^2) \, t_i^{\alpha\mu\nu}(q,r,p) \,, 
\label{eq:GammaXY}
\end{align}
where the explicit expressions of the four transverse tensors $t_i$ and the ten non-transverse tensors $\ell_i$ are given, for instance, in Eqs.~(3.4) and~(3.6) of~\cite{Aguilar:2019jsj}. 

Bose symmetry entails that, after factoring the fully antisymmetric color tensor $f^{abc}$ out of the correlation function, both $\Gamma^{\alpha\mu\nu}(q,r,p)$ 
and $\overline{\Gamma}^{\alpha \mu \nu}(q,r,p)$
reverse their sign under the exchange of Lorentz 
indices and momenta between any two of the incoming gluons, 
\eg \mbox{$\{\alpha, q\} \leftrightarrow \{\mu, r\}$}.  
Since $\overline{\Gamma}^{\alpha \mu \nu}(q,r,p)$
is completely transverse, it could be expanded directly in the basis of tensors $t_i$, which, however, do not individually reverse sign under such an exchange. 
Alternatively, one may construct a basis of 
transverse tensors $\tlambda_i$ according to 
\begin{subequations}
\label{eq:tensors}
\begin{align}
\tlambda_1^{\alpha \mu \nu} \!=& \ P_{\alpha'}^\alpha(q)  P_{\mu'}^{\mu}(r)  P_{\nu'}^\nu(p) 
\left[\ell_1^{\alpha'\mu'\nu'} + \ell_4^{\alpha'\mu'\nu'} + \ell_7^{\alpha'\mu'\nu'}\right] \,,
\label{eq:tl1}
%\nonumber 
\\ 
\tlambda_2^{\alpha \mu \nu} \!=&  \frac{3}{2 s^2} \,(q-r)^{\nu'} (r-p)^{\alpha'} (p-q)^{\mu'} 
P_{\alpha'}^\alpha(q)  P_{\mu'}^{\mu}(r)  P_{\nu'}^\nu(p)\,,
\label{eq:tl2}
%\nonumber
\\
\tlambda_3^{\alpha \mu \nu} \!=& \frac{3}{2 s^2}  P_{\alpha'}^\alpha(q)  P_{\mu'}^{\mu}(r)  P_{\nu'}^\nu(p) 
\left[\ell_3^{\alpha'\mu'\nu'} + \ell_6^{\alpha'\mu'\nu'} + \ell_9^{\alpha'\mu'\nu'}\right]\,,
\label{eq:tl3}
%\nonumber 
\\
\tlambda_4^{\alpha \mu \nu} \!=& \left( \frac{3}{2 s^2}\right)^2
\left[t_1^{\alpha\mu\nu} + t_2^{\alpha\mu\nu} + t_3^{\alpha\mu\nu}\right]\,,
\label{eq:tl4}
\end{align}
\end{subequations}
transforming as $\tlambda_i \to - \tlambda_i$ under a Bose transformation. Employing this latter basis, we have  
\be
\overline{\Gamma}^{\alpha \mu \nu}(q,r,p) = \sum_{i=1}^4 \widetilde{\Gamma}_i(q^2,r^2,p^2) \,
\tlambda_i^{\alpha\mu\nu}(q,r,p) \,, 
\label{eq:expl}
\ee
with the form factors satisfying the special relations 
\be
\widetilde{\Gamma}_i(q^2,r^2,p^2) =\widetilde{\Gamma}_i(r^2,q^2,p^2)
=\widetilde{\Gamma}_i(q^2,p^2,r^2).
\label{eq:brel}
\ee
Thus, in the basis $\{\,\tlambda_i\,\}$, 
Bose symmetry enforces the invariance of the form factors under any exchange of momenta; consequently, the form factors 
must be functions of three Bose-symmetric combinations of $q^2$, $r^2$, and $p^2$, such as the $s^2$ introduced earlier. This particular property, not shared by the corresponding form factors\footnote{Denoted by $\overline{Y}_i$, the form factors of 
$\overline{\Gamma}^{\alpha \mu \nu}(q,r,p)$ 
in the $t_i$ basis obey the constraints\,\cite{Aguilar:2019jsj}: 
$\overline{Y}_1(q,r,p) = \overline{Y}_1(r,q,p)$,\, 
$\overline{Y}_2(q,r,p) = \overline{Y}_2(q,p,r)$,\,
$\overline{Y}_3(q,r,p) = \overline{Y}_3(p,r,q)$,\,
$\overline{Y}_2(q,r,p) = \overline{Y}_1(r,p,q)$, \,
and 
$\overline{Y}_3(q,r,p) = \overline{Y}_1(p,q,r)$.} of the basis $\{\,t_i\,\}$ makes the basis in Eqs.\,\eqref{eq:tensors} especially suitable for our analysis, as will be seen below. Note furthermore that $\tlambda_1$ corresponds to the tree-level case of the transversely projected vertex, a particularly helpful feature when implementing the renormalization prescription. 

The form factors $\widetilde{\Gamma}_i$ may be then projected out from $\overline{\Gamma}^{\alpha\mu\nu}$, 
\begin{equation}\label{eq:Proj3g}
\widetilde{\Gamma}_i(q^2,r^2,p^2) = 
{\cal P}_i^{\alpha\mu\nu}(q,r,p) \overline{\Gamma}_{\alpha\mu\nu}(q,r,p) \;, 
\end{equation}
with the projectors
\begin{equation}\label{eq:Piamn}
 {\cal P}_i^{\alpha\mu\nu}(q,r,p) = \sum_{j=1}^4 \widetilde{M}^{-1}_{ij}(q^2,r^2,p^2) \tlambda_j^{\alpha\mu\nu}(q,r,p) \;,
\end{equation}
defined in terms of the inverse of the $4\times 4$ matrix, whose elements result from the contraction of the basis tensors,
\begin{equation}\label{eq:Mij}
\widetilde{M}_{ij}(q^2,r^2,p^2) = \tlambda_i^{\alpha\mu\nu}(q,r,p) \,\tlambda_{j\ \alpha\mu\nu}(q,r,p) \;. 
\end{equation}
In summary, the transition from the typical two- and three-point lattice correlation functions to the scalar form factors proceeds by first deriving the transversely projected 
three-gluon vertex with the aid of  Eqs.\,(\ref{eq:Green3g},\ref{eq:gGD3},\ref{eq:G2andG3}), and then  employing Eqs.\,(\ref{eq:Proj3g}-\ref{eq:Mij}) to extract from it the corresponding form factors. 

\smallskip

\noindent\textbf{4.$\;$Special kinematics}. 
%\section{Special kinematics}
%
We next focus on the 
bisectoral configurations, $q^2=r^2$ for any $p^2$. 
In that case, 
the subspace spanned by  
$\overline{\Gamma}^{\alpha\mu\nu}$ reduces its dimension down to 3; the determinant of the matrix \eqref{eq:Mij} is therefore vanishing, making the matrix $\widetilde{M}_{ij}$ non-invertible. Indeed, a full tensor basis is obtained by 
\begin{equation}
\lambda_i^{\alpha\mu\nu}(q,r,p) = \lim_{r^2 \to q^2} \tilde{\lambda}_i^{\alpha\mu\nu}(q,r,p) 
\end{equation}    
for i=1,2,3, while, after introducing the dimensionless parameter $z:= p^2/s_b^2$, we have 
\be
\label{eq:lambda4bi}
\lim_{r^2 \to q^2} \lambda_4^{\alpha\mu\nu} =  
\sum_{i=1}^3 f_i(z) \,\lambda_i^{\alpha \mu \nu}(q,r,p)\,, \\ 
\ee
with
\be
f_1(z) = \frac{9}{16} z \,(1-z)\,, \,\,\,\,\,\,\,
f_2(z) = \frac{9}{32} z - \frac{3}{8}\,,\,\,\,\,\,\,\,
f_3(z) = \frac{3}{8} z\,.
\label{eq:thefs}
\ee

Then, we can replace the $4\times 4$ matrix \eqref{eq:Mij} by its $3\times 3$ block in the limit $r^2 \to q^2$, 
\begin{equation}\label{eq:newMij}
\widetilde{M}_{ij}(q^2,r^2,p^2) \Rightarrow M_{ij}(q^2,p^2) = \widetilde{M}_{ij}(q^2,q^2,p^2)
\end{equation}
for $i=1,2,3$; and, after inverting the reshaped matrix, apply the result to Eqs.\,(\ref{eq:Proj3g},\ref{eq:Piamn}) to eventually deliver the three scalar form factors for the bisectoral case. According to Eq.\,\eqref{eq:lambda4bi}, they can be related to the ones in the most general kinematics as

\be
\overline{\Gamma}_i(q^2,q^2,p^2) = \widetilde{\Gamma}_i(q^2,q^2,p^2) + f_i(z) \,\widetilde{\Gamma}_4(q^2,q^2,p^2) \,, 
\label{eq:barGammabi}
\ee
with $i=1,2,3.$

We next consider two kinematic configurations depending 
on a single variable, namely  
the so-called symmetric and soft-gluon limits;  
they are obtained from the bisectoral 
configurations by imposing   
the conditions \mbox{$p^2=q^2$} and \mbox{$p^2=0$}, 
respectively (green and orange points in Fig.~\ref{fig:triangle}) . 
Note that in both cases 
the determinant of the $M_{ij}(q^2,p^2)$ defined in  
Eq.~\eqref{eq:newMij} vanishes; the corresponding ranks are 2 and 1, respectively. 

In the symmetric limit, $\theta_{qr}$=$\theta_{rp}$=$\theta_{pq}$=$2\pi/3$,
the inversion of the $2\times 2$ block of the matrix \eqref{eq:Mij} determines the two basis tensors, whose 
form factors are given by 
%---
\begin{subequations}
\label{eq:gentosym2}
\begin{align}
\overline{\Gamma}_1^{\textrm{sym}}(q^2) &= \lim_{p^2\to q^2} \overline{\Gamma}_1(q^2,q^2,p^2) + \frac 1 2 \,\overline{\Gamma}_3(q^2,q^2,p^2) \;, \\ \label{eq:gentosym2G2}
\overline{\Gamma}_2^{\textrm{sym}}(q^2) &= \lim_{p^2 \to q^2} \overline{\Gamma}_2(q^2,q^2,p^2) - \frac 3 4 \,\overline{\Gamma}_3(q^2,q^2,p^2) \;. 
\end{align}
\end{subequations}  
In the soft-gluon case, $\theta_{qr} \to \pi$; however 
this limit can be approached in different ways, according to $\theta_{rp} \to \theta_l$ and $\theta_{pq} \to \pi - \theta_{l}$; specifically, the 
bisectoral definition corresponds to $\theta_l =\pi/2$.
Notwithstanding this, 
a careful analysis reveals that 
no ambiguity exists\,\cite{Aguilar:2021lke,Aguilar:2021okw},
and that a single form factor emerges, given by 
\begin{equation}\label{eq:gentosg2}
\overline{\Gamma}^{\textrm{sg}}(q^2) = \lim_{p^2 \to 0} \overline{\Gamma}_1(q^2,q^2,p^2) + \frac 3 2 \, \overline{\Gamma}_3(q^2,q^2,p^2)\,.
\end{equation}
Note that the form factors 
$\overline{\Gamma}_{1,2}^{\textrm{sym}}$
and $\overline{\Gamma}^{\textrm{sg}}$ 
have been introduced and evaluated in\,\cite{Aguilar:2021lke,Aguilar:2021okw}.  

Finally, we need to implement multiplicative renormalization by introducing 
the standard renormalization constants, 
%---
\begin{align}
\Delta_R(q^2) &= Z_A^{-1} \Delta(q^2),  &{\cal G}_R(q,r,p) &= Z_A^{-3/2} {\cal G}(q,r,p),  \nonumber 
\\ 
g_R &=  Z_A^{3/2} Z_3^{-1} g, &\overline{\Gamma}_{i\, R}(q^2,q^2,p^2) &= Z_3 \overline{\Gamma}_i(q^2,q^2,p^2)\,,
\label{eq:genren}
\end{align}
%---
relating  bare and renormalized quantities. We specialize here for the \emph{bisectoral} kinematics although the results can be extended to general kinematics. The momentum subtraction (MOM) scheme\,\cite{Hasenfratz:1980kn} is then applied by imposing that renormalized correlation functions must acquire their tree-level expressions at the subtraction point $\mu^2$ (all the renormalized quantities should be understood as depending implicitly on $\mu^2$). 
In the case of the gluon propagator the unique choice is simply 
\mbox{$\Delta^{-1}_{{\rm R}}(\mu^2) = \mu^2$}. Instead, for the  
three-gluon vertex  the renormalization condition must be implemented 
for a specific kinematic configuration; our particular choice is that of the \emph{soft-gluon} kinematics, implying  
%---
\begin{equation}\label{eq:Z3musg}
Z_3^{\mu\textrm{sg}} \overline{\Gamma}_1(\mu^2,\mu^2,0) = 
\overline{\Gamma}_{1\, R}(\mu^2,\mu^2,0) = 1 \;.
\end{equation} 
Thus, projecting out the tree-level component of the three-gluon correlation function, one is left with
\begin{equation}\label{eq:GcalR}
\left. \mathcal{P}_1^{\alpha\mu\nu} {\cal G}_{R\, \alpha\mu\nu}(q,r,p) \right|_{\mu\textrm{sg}} =  g^{\textrm{sg}}_R  \Delta^2_R(\mu^2)  \Delta_R(0) \;,
\end{equation}
with $\mu\textrm{sg}:= \{q^2=r^2=\mu^2,p^2=0\}$. Then, the strong coupling in the \emph{soft-gluon} scheme, $g_R^{\textrm{sg}}=Z_A^{3/2} (Z_3^{\mu\textrm{sg}})^{-1} g$,  remains defined through  Eq.\,\eqref{eq:GcalR}; while $Z_3^{\mu\textrm{sg}}$ is obtained from Eq.\,\eqref{eq:Z3musg} and implies 
\begin{equation}\label{eq:GiR}
\overline{\Gamma}_{i\, R}(q^2,q^2,p^2) = \frac{\overline{\Gamma}_i(q^2,q^2,p^2)}{\overline{ \Gamma}_1(\mu^2,\mu^2,0)} \;, 
\end{equation}
for $i=1,2,3$, owing to multiplicative renormalizability. 

Note that, had we chosen a different kinematic configuration to fix the renormalization condition for the three-gluon vertex as, for instance 
$\mu\text{sym}:=\{q^2=r^2=p^2=\mu^2\}$, we would have been left with
\begin{equation}
Z_3^{\mu\textrm{sym}}  \overline{\Gamma}_1(\mu^2,\mu^2,\mu^2)  = 1 \;;
\end{equation}
and thereby
\begin{equation}
g_R^{\textrm{sym}} = \frac{Z_A^{3/2}}{Z_3^{\mu\textrm{sym}}} \, g =\frac{Z_3^{\mu\textrm{sg}}}{Z_3^{\mu\textrm{sym}}}  \, g_R^{\textrm{sg}} 
=  \frac{ \overline{\Gamma}_1(\mu^2,\mu^2,\mu^2)}{\overline{ \Gamma}_1(\mu^2,\mu^2,0)} \, g_R^{\textrm{sg}}  \,, 
\end{equation}
which relates the coupling in two different schemes.

\smallskip

\noindent\textbf{5.$\;$Vertex form factors from lattice QCD}. 
%\section{Vertex form factors from lattice QCD}
%
In order to obtain lattice results for the three-gluon vertex in extended kinematics, 
we compute the three- and two-points functions of Eqs.\,\eqref{eq:Green3g} and \eqref{eq:Green2g} by sampling Monte-Carlo ensembles of quenched lattice gauge-field configurations produced with the Wilson action (see Tab.\,\ref{tab:setup} for set-up details). The hypercubic  artifacts associated with the breaking of the rotational symmetry O(4) down to H(4) are cured by applying the so-called H4-extrapolation\,\cite{Becirevic:1999uc,Becirevic:1999hj,deSoto:2007ht,deSoto:2022scb} when the number of available H4 orbits and data 
permits it\footnote{ 
This method is known to effectively remove  the hypercubic artifacts in the case of both the propagator and soft-gluon kinematics~\cite{Boucaud:2018xup,Aguilar:2021okw}; therefore, it has been duly applied on them in the present work.}; otherwise, we average over all H4 orbits sharing the same momentum in the continuum limit. Next, we derive the transversely projected 1PI vertex, Eqs.\,\eqref{eq:gGD3} and \eqref{eq:barGamma}, and project out the form factors $\widetilde{\Gamma}_i$ following Eqs.\,(\ref{eq:Proj3g}-\ref{eq:Mij}). Finally, we apply the renormalization prescription given in Eq.\,\eqref{eq:GiR}, 
choosing $\mu$=4.3 GeV as our subtraction point; in what follows, 
the suffix ``$R$'' will be suppressed from all 
renormalized quantities.
All computed errors are purely statistical,  obtained 
through the application of the ``Jack-knife method''. 

\begin{table}[htb]
\begin{center}
	\begin{tabular}{c c c c}
		\hline
		\hline
		$\beta$ & $L^4/a^4$ & a (fm) & confs \\
		\hline
		5.6 & $32^4$  & 0.236 & 2000 \\
		\hline
		5.8 & $32^4$  & 0.144 & 2000 \\
		\hline
		6.0 & $32^4$  & 0.096 & 2000 \\
		\hline
		6.2 & $32^4$  & 0.070 & 2000 \\
		\hline
		\hline
	\end{tabular}
\end{center}	
\caption{Gauge-field configurations (number in the fourth row) produced with the Wilson action, a bare coupling defined by $\beta$ (first row)  	in $L^4$ lattices (second row). The scale setting is made, as described in Ref.\,\cite{Boucaud:2018xup}, by way of a relative calibration based on the scaling of gluon propagators supplemented by the introduction of physical units at $\beta$=5.8 reported in\,\cite{Necco:2001xg} (lattice spacings in the third row).}
	\label{tab:setup}
\end{table}

\begin{figure}[htb]
\begin{tabular}{c}
\includegraphics[width=\columnwidth]{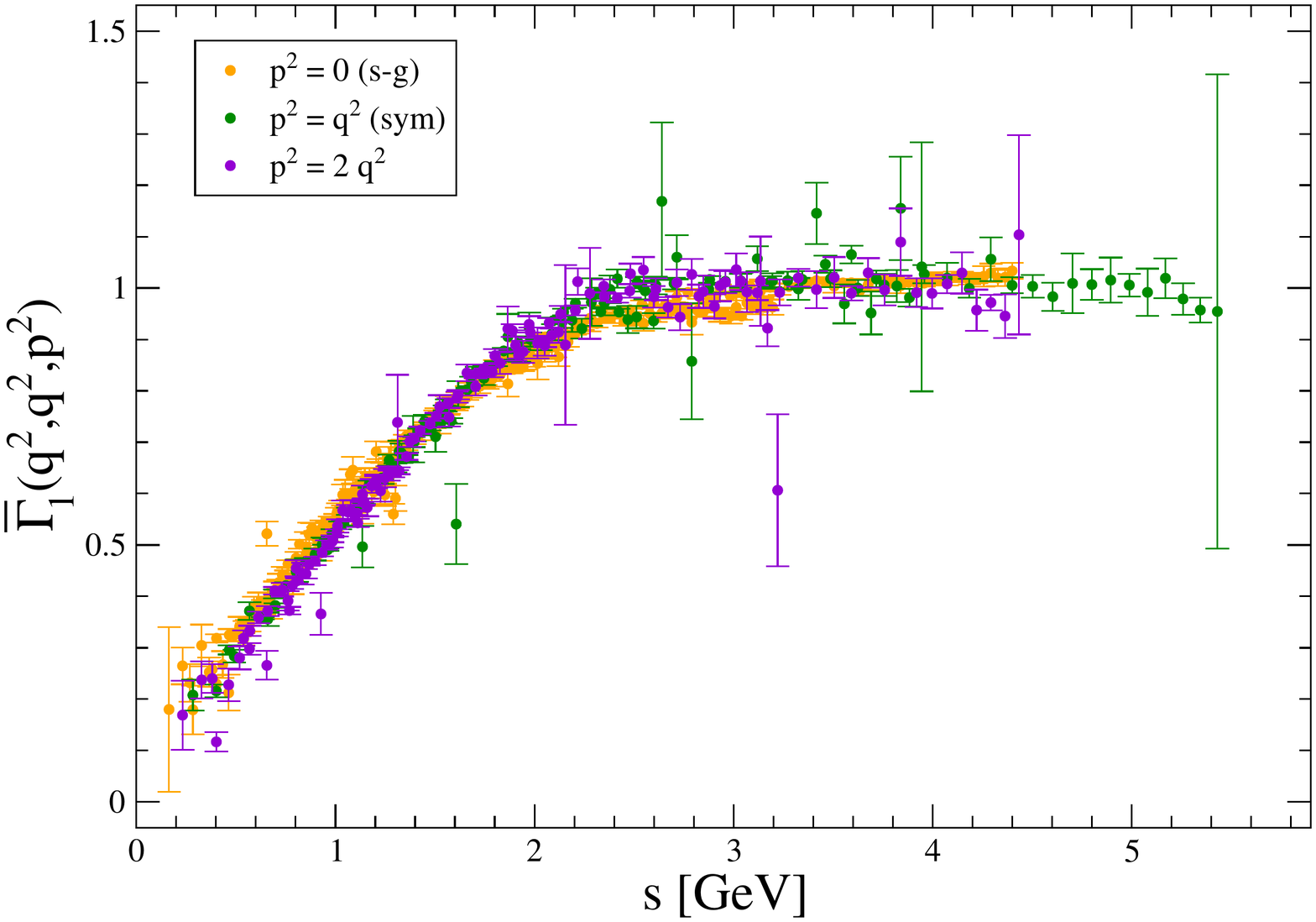} 
\vspace*{-1cm}
\\
\includegraphics[width=\columnwidth]{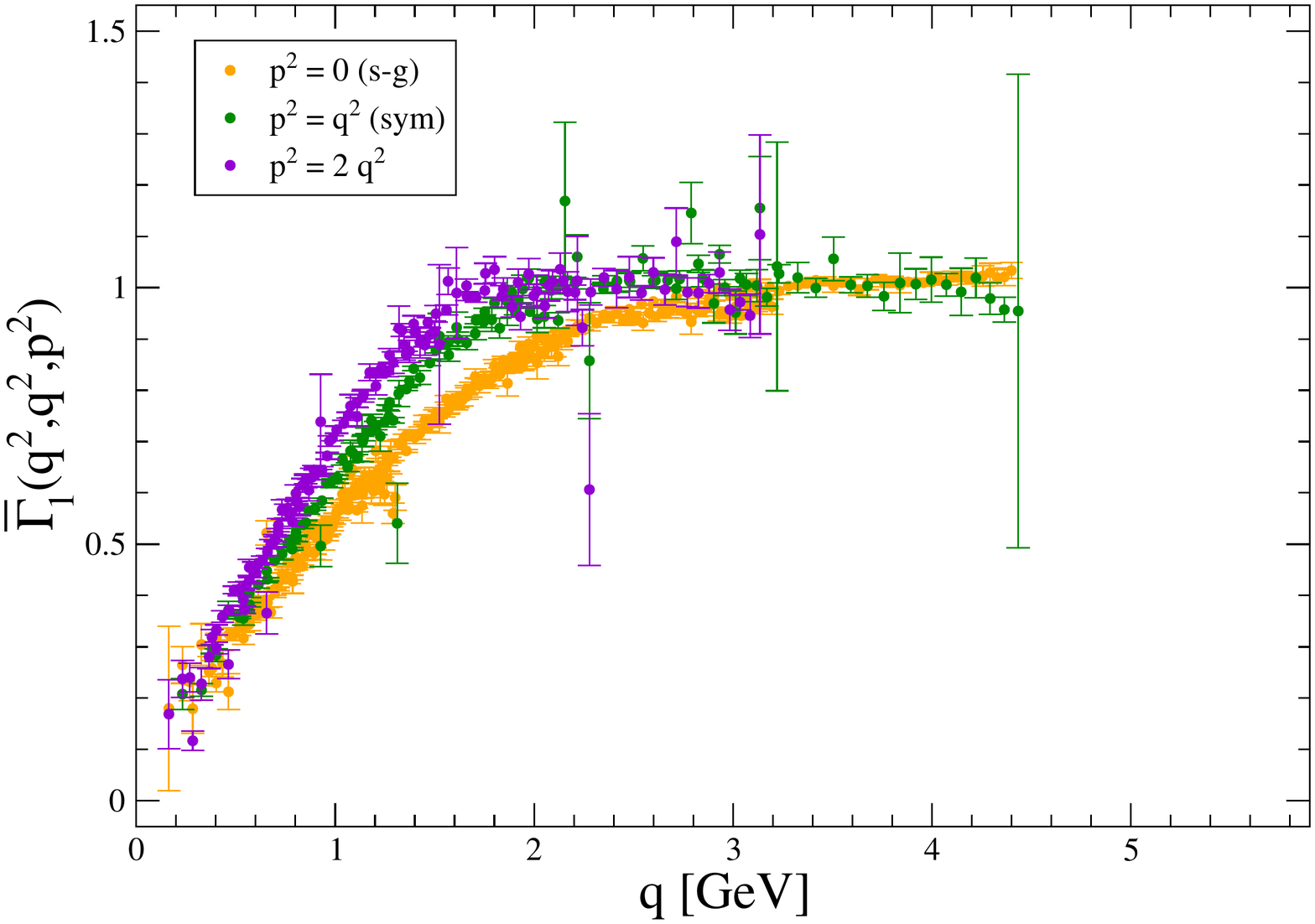}  
\end{tabular}
\vspace*{-0.8cm}
\caption{[Upper plot.-] The scalar form factor $\overline{\Gamma}_{1\, R}$ as a function depending only on the symmetric combination of momenta $s$ for the three classes of configurations displayed in Fig.\,\ref{fig:triangle}: $p^2=0$, $r^2=q^2$ (orange solid circles); $p^2=r^2=q^2$ (green); $p^2=2r^2=2q^2$ (violet). The renormalization point is $\mu$=4.3 GeV. [Bottom.-] Data are displayed also in terms of $q^2$, for the sake of comparison. } 
\label{fig:GammaA3p}
\end{figure}

\begin{figure}[htb]
\includegraphics[width=\columnwidth]{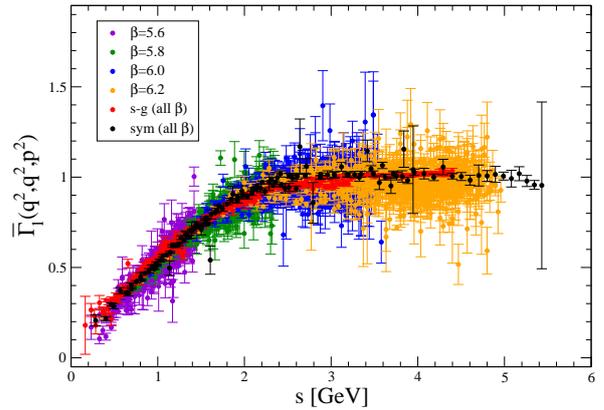} 
\vspace*{-1cm}
\caption{Lattice data obtained from the gauge-field configurations of Tab.\,\ref{tab:setup} for the scalar form factor $\overline{\Gamma}_{1\, R}(q^2,r^2,p^2)$, plotted in terms of $s$. Data for all the kinematic configurations obeying $q^2=r^2$ (lying on the grey line in Fig.\,\ref{fig:triangle}) have been displayed, except the noisiest ones for kinematic configurations very close to symmetric and soft-gluon cases (the projection matrix takes eigenvalues which approach zero). Both cases have been individually treated after reducing the tensor basis, as explained, and the results displayed with black and red solid circles. The renormalization point is $\mu$=4.3 GeV.}
\label{fig:GammaA}
\end{figure}

\begin{figure}[htb]
\vspace*{-0.75cm}
\includegraphics[width=1.1\columnwidth]{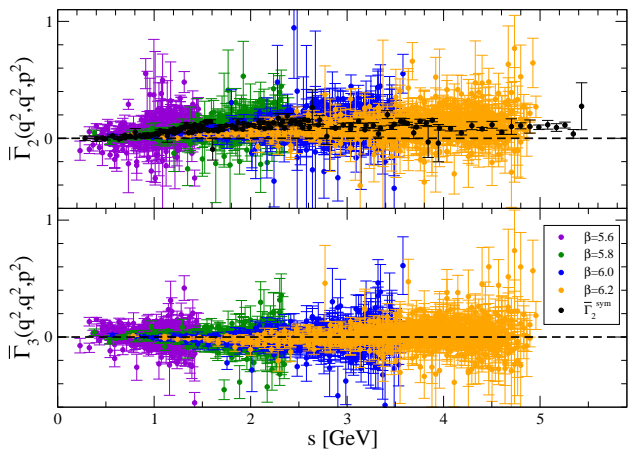} 
\vspace*{-1cm}
\caption{The same as in Fig.\,\ref{fig:GammaA} for $\overline{\Gamma}_2(q^2,r^2,p^2)$ (upper) and $\overline{\Gamma}_3(q^2,r^2,p^2)$ (lower). In the former case the black solid squares stand for the form factor derived in the symmetric limit.}
\label{fig:GammaBandC}
\end{figure}

 Given our choice of tensor basis, the scalar form factors can only depend on three Bose-symmetric combinations of momenta. 
The most obvious is $s^2 = \frac 1 2 \left( q^2 + r^2 + p^2 \right)$, obtained directly  from the rotated coordinate in \1eq{eq:hatp}; 
two more 
can be built, for instance, through the symmetrization of \2eqs{eq:hatq}{eq:hatr}. Although all three variables are in principle required 
for a full kinematic description of the vertex form factors, 
our lattice results indicate that only $s^2$ is relevant, 
while the dependence on the two others is severely suppressed. 
Note that this particular results are consistent with the findings of the study presented in \,\cite{Eichmann:2014xya}. 

According to the above observation, 
all kinematic configurations lying on a plane $\hat{p}^2$=cte. share the same form factors.
This is rather striking, because,  
\emph{a priori}, Bose symmetry alone can only 
enforce the equality between the form factors 
of the few kinematic configurations that are connected by simple permutations of momenta. 
The degree of validity of this exceptional property 
is illustrated in Figs.\,\ref{fig:GammaA3p}-\ref{fig:GammaBandC} 
for the form factors of the bisectoral kinematics.

In particular, 
Fig.\,\ref{fig:GammaA3p} displays $\overline{\Gamma}_1$ in the three kinematic configurations highlighted in Fig.\,\ref{fig:triangle}: the agreement at equal $s^2$ is 
contrasted to the considerable disparity seen at equal $q^2$. 
Even though fairly apparent 
to the naked eye, the coincidence of values achieved when 
using $s^2$ instead of $q^2$ is 
even more impressive  when expressed in numbers. 
Specifically, 
given two sets of data $\{x_i,y_i,\delta_i(y)\}$ and $\{x_i,z_i,\delta_i(z)\}$, sharing $N$ common values of $x$, one may define $\chi^2/\textrm{datum}=N^{-1} \sum_{i=1}^N (y_i-z_i)^2/(\delta_i^2(y)+\delta_i^2(z))^2$. This definition 
allows us to measure the "dispersion" of the symmetric and $p^2$=$2q^2$ 
configurations from the soft-gluon data, which act as our reference set. 
Initially, this computation yields two pairs
of numbers; then, the elements of each pair are weighted 
according to the number of points contained in each set, and 
finally averaged. The final result is 
$\chi^2/\textrm{datum}$=66.4 for the data in the lower panel of Fig.\ref{fig:GammaA3p}, and $\chi^2/\textrm{datum}$ =3.9
for those in the upper. Thus, with the aid of this procedure, we 
conclude that 
plotting the data as a function of $s^2$ increases the 
"overlap" between the curves by a factor of 17. 

Quite 
remarkably, this pattern  persists for  
{\it all} bisectoral configurations shown in 
Fig.\,\ref{fig:GammaA}, where approximately  
4000 data points are included.
Furthermore, the form factor $\overline{\Gamma}_2$, depicted in the upper panel of  Fig.\,\ref{fig:GammaBandC}, although amounting to about one tenth of $\overline{\Gamma}_1$, also depends solely on $s^2$. 
Finally, the $\overline{\Gamma}_3$ shown in the  
lower panel is compatible with zero within the errors. 

In addition, even though a systematic
exploration is pending, a random scanning of 
the allowed kinematic region beyond $q^2$=$r^2$ 
confirms the above results 
for $\widetilde{\Gamma}_i$, with $i$=1,2,3, while 
$\widetilde{\Gamma}_4$ remains negligible. The latter indicates that, according to Eqs.\,\eqref{eq:barGammabi}, $\widetilde{\Gamma}_i \approx  \overline{\Gamma}_i$. All the above findings justify the following 
approximate relations
%---
\begin{subequations}
\begin{align} 
\label{eq:Gamma1allkin}
\widetilde{\Gamma}_1(q^2,r^2,p^2)% \equiv \overline{\Gamma}_{1\, R}(s^2) 
&\approx \overline{\Gamma}_1(s^2,s^2,0) \approx \overline{\Gamma}^{\textrm{sg}}(s^2) \;, \\   
\label{eq:Gamma2allkin}
\overline{\Gamma}_2(q^2,r^2,p^2) %\equiv \overline{\Gamma}_{2\, R}(s^2) 
&\approx \widetilde{\Gamma}_2\left(\frac{2s^2}{3},\frac{2s^2}{3},\frac{2s^2}{3}\right) \approx  \overline{\Gamma}_2^{\textrm{sym}}\left(\frac{2s^2}{3}\right) \;;  \end{align}
\end{subequations}
where $\overline{\Gamma}^{\textrm{sg}}$ and $\overline{\Gamma}^{\textrm{sym}}_2$ denote  the soft-gluon and symmetric form factors given in Eqs.\,\eqref{eq:gentosg2} and \eqref{eq:gentosym2G2}, respectively. 
Armed with these results, we can derive the key relation  
\begin{equation}
\label{eq:Gmnaallkin}
\overline{\Gamma}^{\alpha\mu\nu}(q,r,p) = \overline{\Gamma}^{\textrm{sg}}\left(s^2 \right) \widetilde{\lambda}_1^{\alpha\mu\nu}(q,r,p) + \overline{\Gamma}_2^{\textrm{sym}}\left(\frac{2s^2}{3}\right) \widetilde{\lambda}_2^{\alpha\mu\nu}(q,r,p)\,,
\end{equation}
which serves 
as an excellent approximation for the transversely projected three-gluon vertex. 

The form factors $\overline{\Gamma}^{\textrm{sg}}$ and $\overline{\Gamma}^{\textrm{sym}}_2$ 
have been studied in detail in a series of recent articles\,\cite{Aguilar:2019uob,Aguilar:2021okw,Aguilar:2021lke}.
In fact, due to the established dominance of $\overline{\Gamma}^{\textrm{sg}}$ over $\overline{\Gamma}^{\textrm{sym}}_2$, 
\1eq{eq:Gmnaallkin} may be further simplified to 
\begin{equation}
\label{eq:compact}
\overline{\Gamma}^{\alpha\mu\nu}(q,r,p) \approx \overline{\Gamma}^{\textrm{sg}}\left(s^2 \right) \widetilde{\lambda}_1^{\alpha\mu\nu}(q,r,p) \,,
\end{equation} 
providing an exceptionally compact approximation 
for $\overline{\Gamma}^{\alpha\mu\nu}(q,r,p)$ in general kinematics.

\begin{figure}[htb]
\centering
\begin{tabular}{c}
\includegraphics[width=0.8\columnwidth]{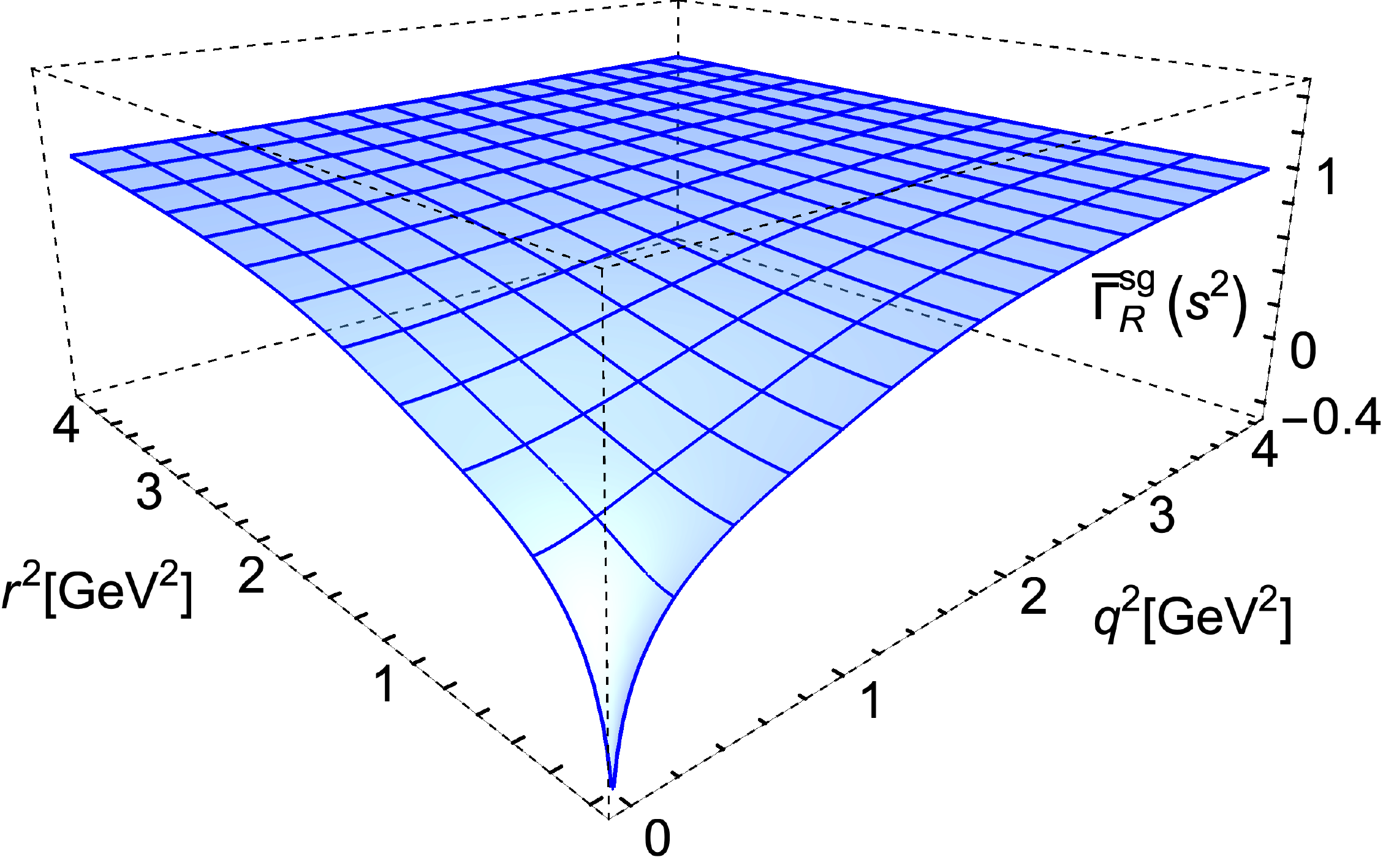} \\
\includegraphics[width=0.8\columnwidth]{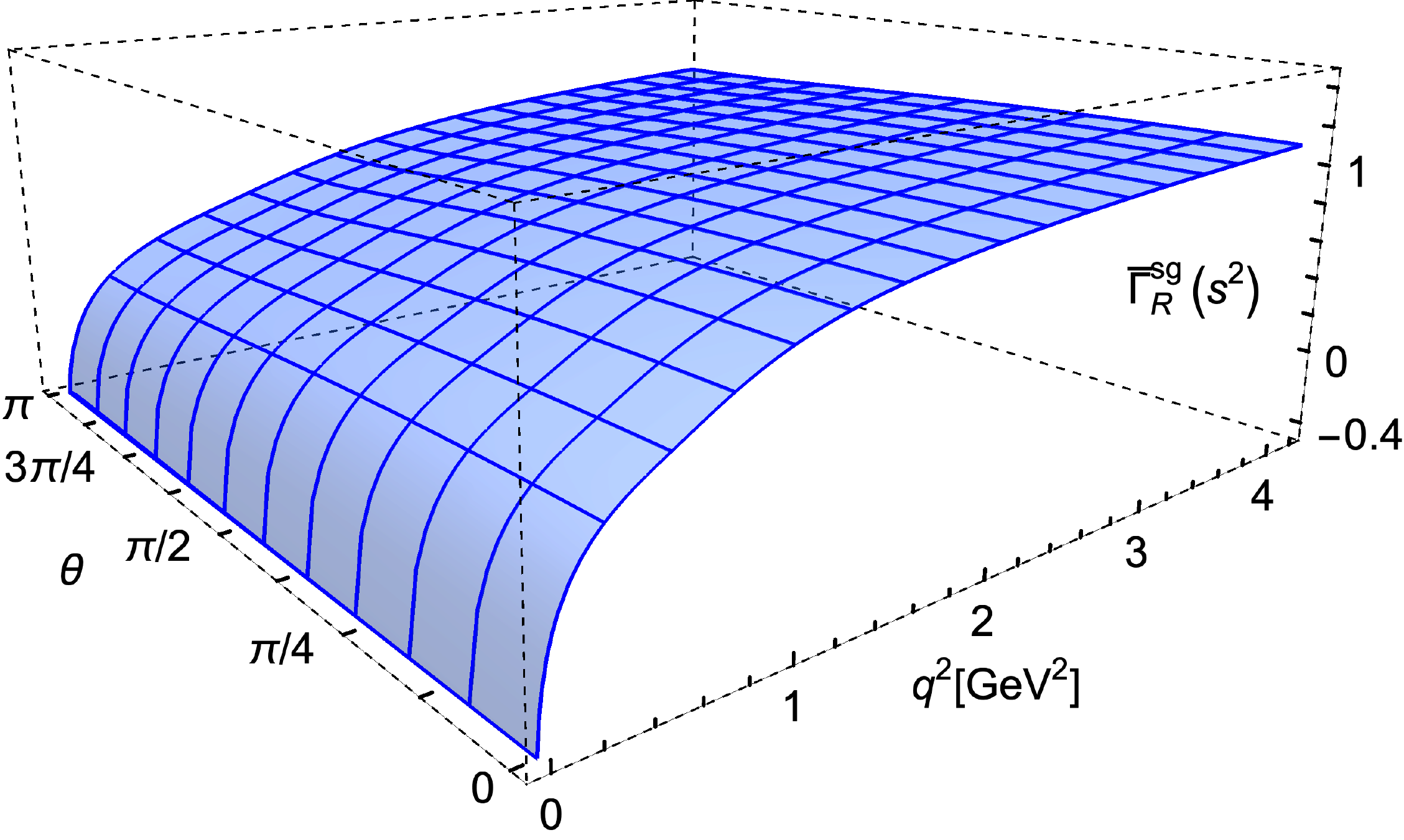}
\end{tabular}
\vspace*{-0.25cm}
\caption{The form factor $\overline{\Gamma}^{\textrm{sg}}(s^2)$, displayed in terms of $q^2$ and $r^2$, with $p^2$ fixed by $\cos{\theta_{qr}}=\pi/2$ (upper panel); and in terms of $q^2$ and $\theta_{qr}$ for $q^2$
=$r^2$ and any allowed $p^2$ (lower panel).}
\label{fig:3d}
\end{figure}

Given that $\overline{\Gamma}^{\textrm{sg}}$ 
is the sole dynamical ingredient in \1eq{eq:compact}, 
we depict it in Fig.\,\ref{fig:3d}; 
there one may appreciate the characteristic 
logarithmic divergence at the origin, induced by 
the ghost loops (upper panel), 
and the mild dependence on the angle $\theta$  (lower panel).  

\smallskip

\begin{figure}[htb]
\includegraphics[width=\columnwidth]{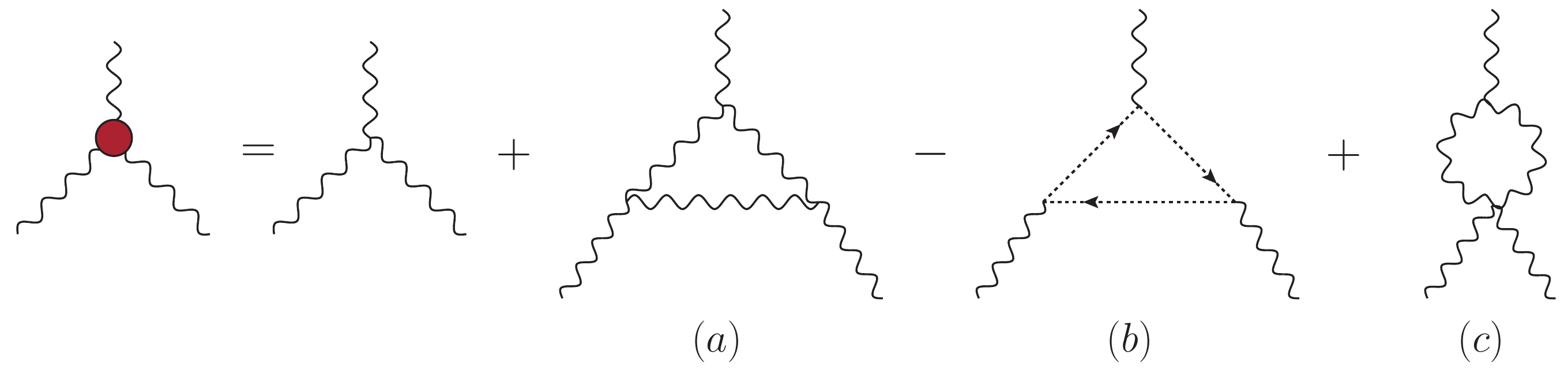}
\vspace*{-0.5cm}
\caption{Feynman diagrams contributing to the three-gluon vertex at one-loop; crossed diagrams are not shown.}
\label{fig:1loop}
\end{figure}

\noindent\textbf{6. One-loop analysis with a gluon mass}. 
In order to gain some basic insights on the 
origin of the planar degeneracy of $\overline{\Gamma}_1(q^2,r^2,p^2)$, 
we compute this particular 
form factor at the one-loop level, through the diagrams shown in Fig.~\ref{fig:1loop}. In the bisectoral configurations, where $p^2 = 2 q^2( 1 + \cos\theta)$ and $s^2 = q^2( 2 + \cos\theta)$, with $\theta:=\theta_{rq}$, the result can be expressed in the form
\begin{align} 
\overline{\Gamma}_1(q^2,q^2,p^2) =& 1 + w_1 \ln \left(\frac{s^2}{\mu^2}\right) + w_2 \ln{\left( 1+ \cos \theta \right)} + w_3 \,, 
\end{align}
where the $w_i$ are functions of $s^2$ and $\theta$. 
It turns out that the $w_i$ are nonvanishing both at 
$s^2 = 0$ and at $\theta = \pi$ ($p^2 = 0$); consequently, $\overline{\Gamma}_1$ diverges logarithmically in these two limits\footnote{Due to the divergence of the perturbative $\overline{\Gamma}_1$ at $p^2 = 0$, we apply the renormalization condition \1eq{eq:Z3musg} to $\overline{\Gamma}_{\textrm{sg}}=\overline{\Gamma}_1 + 3/2 \overline{\Gamma}_3$, which is found to be finite, instead of $\overline{\Gamma}_1$. Applied nonperturbatively to the lattice form factors, both prescriptions are equivalent as long as $\overline{\Gamma}_3$ is found to be compatible with zero (Fig.\,\ref{fig:GammaBandC}).}.
Note, however, that the nonperturbative fate 
of these two divergences is 
entirely different: while the former is intimately related with 
the masslessness of the ghost, a feature that 
persist nonperturbatively\,\cite{Alkofer:2000wg,Fischer:2006ub,Aguilar:2008xm,Boucaud:2008ky,Boucaud:2008ji}, 
the latter disappears  
when the bona-fide nonperturbative behavior of the 
gluon propagator\,\cite{Aguilar:2008xm,Boucaud:2008ky,Fischer:2008uz,Dudal:2008sp,Papavassiliou:2022wrb}, characterized by the emergence of a gluon mass, 
is taken minimally into account. 

To study this last point in some detail, let us first 
point out that 
the divergence at 
$\theta = \pi$ implies a pronounced deviation from the planar degeneracy 
observed on the lattice; indeed, as  
$\theta$ approaches $\pi$, 
$\overline{\Gamma}_1(q^2,q^2,p^2)$ cannot possibly 
depend on $s^2$ alone. 

To quantify the deviation from the planar degeneracy  
within different computational frameworks, we introduce the function
\begin{align}
d(s^2,\theta) =& \left[\overline{\Gamma}_1(q^2,q^2,p^2) - \overline{\Gamma}_{\rm sg}(s^2) \right]/\overline{\Gamma}_{\textrm{sg}}(s^2) \,, \label{di_def}
\end{align}
which, in the case of exact planar degeneracy, vanishes  
for every value of $\theta$. 
In what follows we evaluate $d(s^2,\theta)$ in three different ways:
({\it i)} calculating the one-loop diagrams of Fig.~\ref{fig:1loop};
({\it ii}) calculating the same diagrams as in ({\it i}), but 
using massive gluon propagators, \ie $\Delta(q^2)\to (q^2+m^2)^{-1}$, 
with $m = 350$ MeV, a value motivated by general theoretical results (\emph{e.g.}, see Ref.\,\cite{Papavassiliou:2022wrb});
and ({\it iii}) using the lattice data  displayed in Fig.\,\ref{fig:GammaA}.
The results of these three cases are summarized in Fig.~\ref{fig:d_pert},
where they are depicted as functions of the angle $\theta$.
Specifically: 

\begin{figure}[t!]
\vspace*{-0.75cm}
\includegraphics[width=\columnwidth]{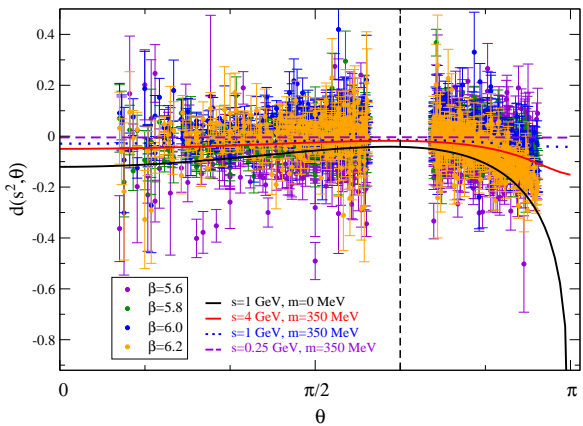}
\vspace*{-0.75cm}
\caption{ The function $d(s^2,\theta)$, defined in \1eq{di_def}, for fixed values of $s^2$ from one-loop calculations and for the available momentum range from the lattice data of Fig.\,\ref{fig:GammaA}. The black solid curve shows the standard perturbative result for $s^2 = 1$ GeV. The results with a massive gluon are shown as violet dashed, blue dotted and red solid curves, for $s = 0.25$ GeV, $s = 1$ GeV and $s = 4$ GeV, respectively. The evaluations were done using $\alpha_s = 0.27$, $\mu = 4.3$ GeV and $m = 350$ MeV for the gluon mass.}
\label{fig:d_pert}
\end{figure}

({\it i}) The $d(s^2,\theta)$ corresponding to the one-loop calculation, 
for the representative value of $s = 1$ GeV, 
is displayed  as a black continuous curve; 
evidently, $d(s^2,\theta)$ is small for \mbox{$\theta \lesssim 3\pi/4$}, but diverges as $\theta \to\pi$.

({\it ii}) The $d(s^2,\theta)$ obtained using massive gluons 
are shown as violet dashed, blue dotted, and red solid curves, for $s = 0.25$ GeV, $s = 1$ GeV and $s = 4$ GeV, respectively. 
Clearly, the inclusion of a gluon mass 
not only makes $d(s^2,\theta)$ finite at $\theta = \pi$, but 
also reduces its overall (absolute) size relative to its massless 
counterpart of case  ({\it i}). 
We note that the point-by-point deviation from the 
planar degeneracy grows as $s$ increases. 
However, 
within the range of $s$ that we have considered, the 
deviation remains below $5\%$, except 
in the vicinity of $\theta=\pi$, where it peaks at $16\%$ for 
$s = 4$ GeV.

({\it iii}) The $d(s^2,\theta)$ extracted from the lattice data, as in Figs.\,\ref{fig:GammaA},\ref{fig:GammaBandC}, is displayed with a given color identifying data from the same $\beta$ (see Tab.\,\ref{tab:setup}), each covering a different domain of momenta: data for $\beta$=5.6 (violet) represent momenta below 1.5 GeV, $\beta$=5.8 (green) below 2.5 GeV, $\beta$=6.0 (blue) below 3.7 GeV, and $\beta$=6.2 (orange) below 5 GeV. Data around $\theta$=$2\pi/3$ are plagued by large statistical noise, and have been excluded. Note that the deviation from planar degeneracy is compatible with zero within the errors, except in the vicinity of $\theta$=$\pi$, where data from ensembles reaching larger momenta are seen to deviate more, following the same tendency displayed in case ({\it ii}), and remaining compatible with the calculations with massive gluons. In fact, lattice data near $\theta$=$\pi$ clearly differ from the result with massless gluons, the bulk of data in this region lying significantly above the curve. We emphasize that the bisectoral kinematic domain has been sampled with, approximately, 4000 data, 700 of which are located within the interval $\theta \in (5\pi/6,\pi)$, thus 
furnishing ample statistical validation to the above statements.

This analysis suggests that the approximate planar degeneracy of $\overline{\Gamma}_1$  
may hinge on the emergence of a mass scale in the gauge sector 
of QCD, as described in the many works cited earlier.

\noindent\textbf{7.$\;$Conclusions}. 
%\section{Conclusions}
%
In the present lattice study we have explored 
the transversely projected three-gluon vertex for 
general kinematics, with particular emphasis on the 
bisectoral configurations, defined by the condition $q^2 = r^2$.
When expressed in a Bose-symmetric 
basis, the vertex form factors depend predominantly on the special 
variable $s^2$; the simple geometric interpretation 
of this property leads to the notion of "planar degeneracy".

In addition, if the tensor basis is chosen such that the tree-level tensor is one of its elements, the corresponding 
form factor clearly dominates over the others. This allows for a  simplified representation of the vertex in terms of the tree-level tensor and a single form factor, whose extended kinematic behavior can be reduced to that of the soft-gluon case, as captured by 
\2eqs{eq:Gmnaallkin}{eq:compact}. 

A systematic lattice study exploring  
the entire kinematic domain is currently underway;   
the preliminary analysis indicates   
that the planar degeneracy persists, 
at a high level of accuracy, beyond the bisectoral configurations.  
The conclusive confirmation of these findings would induce  
vast simplifications to a number of physical applications 
that depend on the detailed knowledge of the three-gluon vertex.
In such cases, the use of formulas
\2eqs{eq:Gmnaallkin}{eq:compact}
could reduce substantially the numerical effort required. 

Let us finally emphasize that, at present, the near "planar degeneracy"
is an empirical observation corroborated by a large number 
of lattice data and a one-loop calculation minimally supplemented by a gluon mass. However, no deeper understanding of its origin is available to us. It would be clearly important to 
unravel the mechanism underlying this particular property and  
establish possible connections with other fundamental aspects of QCD. 

\smallskip

\noindent\textbf{Acknowledgments}. 
The authors thank A. C. Aguilar, G. Eichmann and C. D. Roberts for useful discussions. M.~N.~F.  acknowledges financial support from  the FAPESP projects 2017/05685-2 and 2020/12795-1, respectively. J.~P. is supported by the Spanish MICINN grant PID2020-113334GB-I00 
and the regional Prometeo/2019/087 from the Generalitat Valenciana; while  
F.~D.~S. and J.~R.~Q.~ are by the Spanish PID2019-107844-GB-C2 and the regional Andalusian P18-FR-5057.
All calculations have been performed at the UPO computing center, C3UPO.

\vspace*{-0.25cm}

\bibliographystyle{model1a-num-names}
\bibliography{bibliography}

\end{document}